\documentclass[reprint,twocolumn,citeautoscript,showpacs,amsmath,amssymb,prb]{revtex4}
\usepackage{graphicx}
\usepackage{dcolumn}
\usepackage{bm}
\usepackage{times}
\begin{document}
\title{Magnetic ordering and structural distortion in Ru doped BaFe$_2$As$_2$ single crystals studied by Neutron and X-ray diffraction}
\author{M. G. Kim, D. K. Pratt, G. E. Rustan, W. Tian, J. L. Zarestky, A. Thaler, S. L. Bud'ko, P. C. Canfield, R. J. McQueeney, A. Kreyssig, and A. I. Goldman}
\affiliation{\\Ames Laboratory, U.S. DOE and Department of Physics and Astronomy\\
Iowa State University, Ames, IA 50011, USA}

\date{\today}


\begin{abstract}
We present a systematic investigation of the antiferromagnetic
ordering and structural distortion for the series of
Ba(Fe$_{1-x}$Ru$_x$)$_2$As$_2$ compounds ($0 \leq x \leq 0.246$).
Neutron and x-ray diffraction measurements demonstrate that, unlike
for the electron-doped compounds, the structural and magnetic
transitions remain coincident in temperature.  Both the magnetic and
structural transitions are gradually suppressed with increased Ru
concentration and coexist with superconductivity.  For samples that
are superconducting, we find strong competition between
superconductivity, the antiferromagnetic ordering, and the
structural distortion.
\end{abstract}

\maketitle
\section{Introduction}
After the discovery of FeAs based
superconductors\cite{kamihara_iron-based_2008,rotter_superconductivity_2008},
extensive studies using neutron and x-ray scattering techniques have
revealed strong and unusual interconnections between structure,
magnetism, and superconductivity. In the undoped parent compounds of
the $AE$Fe$_2$As$_2$ ($AE$ = Ba, Sr, Ca) family, the
tetragonal-to-orthorhombic transition and the
paramagnetic-to-antiferromagnetic transition occur at the same
temperature, implying a strong coupling between structure and
magnetism\cite{rotter_spin-density-wave_2008,huang_neutron-diffraction_2008,jesche_strong_2008,goldman_lattice_2008}.
Upon hole-doping with K on the Ba site or electron-doping with
transition metals (e.g. Co, Ni, Rh, Pt, Pd) on the Fe site, the
structural transition temperature ($T_S$) and the antiferromagnetic
(AFM) transition temperature ($T_N$) are suppressed to lower
temperatures.\cite{rotter_superconductivity_2008,chen_coexistence_2009,inosov_suppression_2009,saha_superconductivity_2010,sefat_superconductivity_2008,ni_effects_2008,lester_2009,li_superconductivity_2009,ni_phase_2009,harriger_transition_2009,pratt_coexistence_2009,kreyssig_suppression_2010,wang_electron-doping_2010}
The structural and AFM transitions split with $T_S$ $>$ $T_N$ in
transition-metal doped BaFe$_2$As$_2$,
\cite{sefat_superconductivity_2008,ni_effects_2008,li_superconductivity_2009,ni_phase_2009,harriger_transition_2009,pratt_coexistence_2009,kreyssig_suppression_2010,wang_electron-doping_2010}
whereas the transitions remain coincident in K doped
BaFe$_2$As$_2$.\cite{rotter_superconductivity_2008,chen_coexistence_2009,inosov_suppression_2009}
When both the structural and magnetic transitions are suppressed to
sufficiently low temperatures, independent of the coincidence of
$T_S$ and $T_N$, superconductivity emerges and coexists with
antiferromagnetism for some doping
levels.\cite{pratt_coexistence_2009,kreyssig_suppression_2010,wang_electron-doping_2010}
Moreover, in Co-, Rh-, and Ni-doped BaFe$_2$As$_2$, several neutron
measurements manifest a distinctive suppression of the magnetic
order parameter in the superconducting regime, which clearly
indicates competition between AFM and
superconductivity.\cite{pratt_coexistence_2009,kreyssig_suppression_2010,wang_electron-doping_2010}
Additionally, high-resolution x-ray diffraction measurements on Co-
and Rh-doped BaFe$_2$As$_2$ have revealed the suppression of
orthorhombic distortion below $T_c$ illustrating an unusual
magnetoelastic coupling in the form of emergent nematic order in the
iron
arsenides.\cite{kreyssig_suppression_2010,nandi_anomalous_2010,fang_theory_2008,xu_ising_2008}

In stark contrast to the doping studies mentioned above, hole-doping
through the substitution of
Cr\cite{Sefat_2009,Budko_2009,Marty_2010} or
Mn\cite{Kim_2010,Liu_2010,mgkim_mn_2010} on the Fe site results in
very different behavior.  Neither Ba(Fe$_{1-x}$Cr$_x$)$_2$As$_2$ nor
Ba(Fe$_{1-x}$Mn$_x$)$_2$As$_2$ are superconducting at ambient
pressure for any $x$ and the suppression of the AFM order with
increasing $x$ is more gradual than for the electron-doped series.
Furthermore, for Ba(Fe$_{1-x}$Cr$_x$)$_2$As$_2$ the structural and
magnetic transitions remain locked together up to $x \approx 0.30$
where the stripe-like AFM structure is replaced by G-type AFM order
as found for BaMn$_2$As$_2$\cite{YSingh_2009} and proposed for
BaCr$_2$As$_2$.\cite{DSingh_2009}  For
Ba(Fe$_{1-x}$Mn$_x$)$_2$As$_2$, the structural and AFM transitions
remain locked together until $x \geq 0.102$, where the orthorhombic
distortion abruptly vanishes.\cite{mgkim_mn_2010} We have previously
proposed that, in the absence of the orthorhombic distortion, the
AFM structure may be described by a two-\textbf{Q}
ordering.\cite{mgkim_mn_2010}

Whereas all of the studies above describe measurements performed on
either electron-doped or hole-doped materials, it is also important
to consider the response of these systems to isoelectronic doping.
For example, superconductivity is observed with a maximum $T_c$
$\sim$ 30 K by the isoelectronic doping of P at the As site in
BaFe$_2$As$_2$.\cite{jiang_superconductivity_2009}  Furthermore,
Klintberg \emph{et al}.\cite{Klintberg_2010} have discussed the
equivalence of chemical and physical pressure in
BaFe$_2$(As$_{1-x}$P$_x$)$_2$ by showing that the
temperature-pressure phase diagrams are similar, but shifted for
different $x$.  Nevertheless, the maximum superconducting transition
temperatures are identical. It is believed that superconductivity in
this compound originates from steric effects arising from the
smaller ionic radius of P. Only small modifications of the Fermi
surface were observed.\cite{shishido_evolution_2010}
Superconductivity has also been reported in
Sr(Fe$_{1-x}$Ru$_x$)$_2$As$_2$ compounds with $T_c$ up to 20 K, but
at much higher doping levels than required for the electron-doped
series (e.g. Co, Ni, Rh).\cite{Schnelle_2009,Qi_2009} Ru
substitution on the Fe site in Ba(Fe$_{1-x}$Ru$_x$)$_2$As$_2$ was
recently reported to exhibit properties similar to the
electron-doped BaFe$_2$As$_2$ series but, again, at higher doping
compositions.\cite{sharma_superconductivity_2010,rullier-albenque_hole_2010,thaler_physical_2010,hodovanets_2010}
The structural and AFM transition temperatures are suppressed with
increasing $x$ and superconductivity occurs at $x \approx 0.16$.

Thaler \emph{et al}.~\cite{thaler_physical_2010} have made an
interesting comparison between the phase diagrams of Ru-doped
BaFe$_2$As$_2$ and the parent BaFe$_2$As$_2$ compound under
pressure. Although the unit cell volume increases with Ru doping,
they found a striking similarity between the phase diagrams for Ru
doping and physical pressure when scaled by the lattice parameter
$c$/$a$ ratio.  Only a single feature corresponding to a magnetic,
structural, or joint magnetic/structural transition has been
observed in resistance and magnetization data for
Ba(Fe$_{1-x}$Ru$_x$)$_2$As$_2$ ($x \leq 0.37$), similar to what has
been found for the nonsuperconducting hole-doped series, but quite
different from the behavior of electron doped BaFe$_2$As$_2$.
Interestingly, we note that in the case of P doping on the As site,
a splitting between the structural and magnetic transitions was
noted in resistance measurements, that increases with P
concentration.\cite{Kasahara_2010} It is, therefore, particularly
important to clarify the microscopic nature of the magnetic and/or
structural transitions for the case of isoelectronic doping on the
Fe site in Ba(Fe$_{1-x}$Ru$_x$)$_2$As$_2$, as well as the
interaction between magnetism, structure and superconductivity in
this series.

Here we report on magnetic neutron diffraction and high-resolution
x-ray diffraction measurements on the series of
Ba(Fe$_{1-x}$Ru$_x$)$_2$As$_2$ compounds ($0 \leq x \leq 0.246$)
which demonstrate that, unlike the electron-doped compounds, the
structural and magnetic transitions remain coincident in
temperature.  Similar to the electron-doped samples, however, we
find strong competition between superconductivity, the AFM ordering
and the structural distortion. The transition temperatures,
magnitudes of the ordered magnetic moment, and the magnitude of the
orthorhombic distortions in Ba(Fe$_{1-x}$Ru$_x$)$_2$As$_2$ are
compared with previous reports on Ba(Fe$_{1-x}$Co$_x$)$_2$As$_2$ and
Ba(Fe$_{1-x}$Mn$_x$)$_2$As$_2$~\cite{nandi_anomalous_2010,fernandes_unconventional_2010,mgkim_mn_2010}.

\section{Experiment}
Single crystals of Ba(Fe$_{1-x}$Ru$_x$)$_2$As$_2$ were grown out of
a FeAs self-flux using conventional high temperature solution growth
technique described in Ref.~\onlinecite{thaler_physical_2010}. The
compositions were measured at between 10 and 20 positions on samples
from each growth batch using wavelength dispersive spectroscopy
(WDS). The combined statistical and systematic error on the Ru
composition is not greater than 5\% (e.g. 0.126$\pm$0.003, see
Ref.~\onlinecite{thaler_physical_2010}). Magnetization and
temperature-dependent AC electrical resistance data ($f$ = 16 Hz,
$I$ = 3 mA) were collected in a Quantum Design Magnetic Properties
Measurement System using a Linear Research LR700 resistance bridge
for the latter. Electrical contact was made to the sample using
Epotek H20E silver epoxy to attach Pt wires in a four-probe
configuration.

\begin{figure}[t!]
\begin{center}
\includegraphics[width=1.0\linewidth]{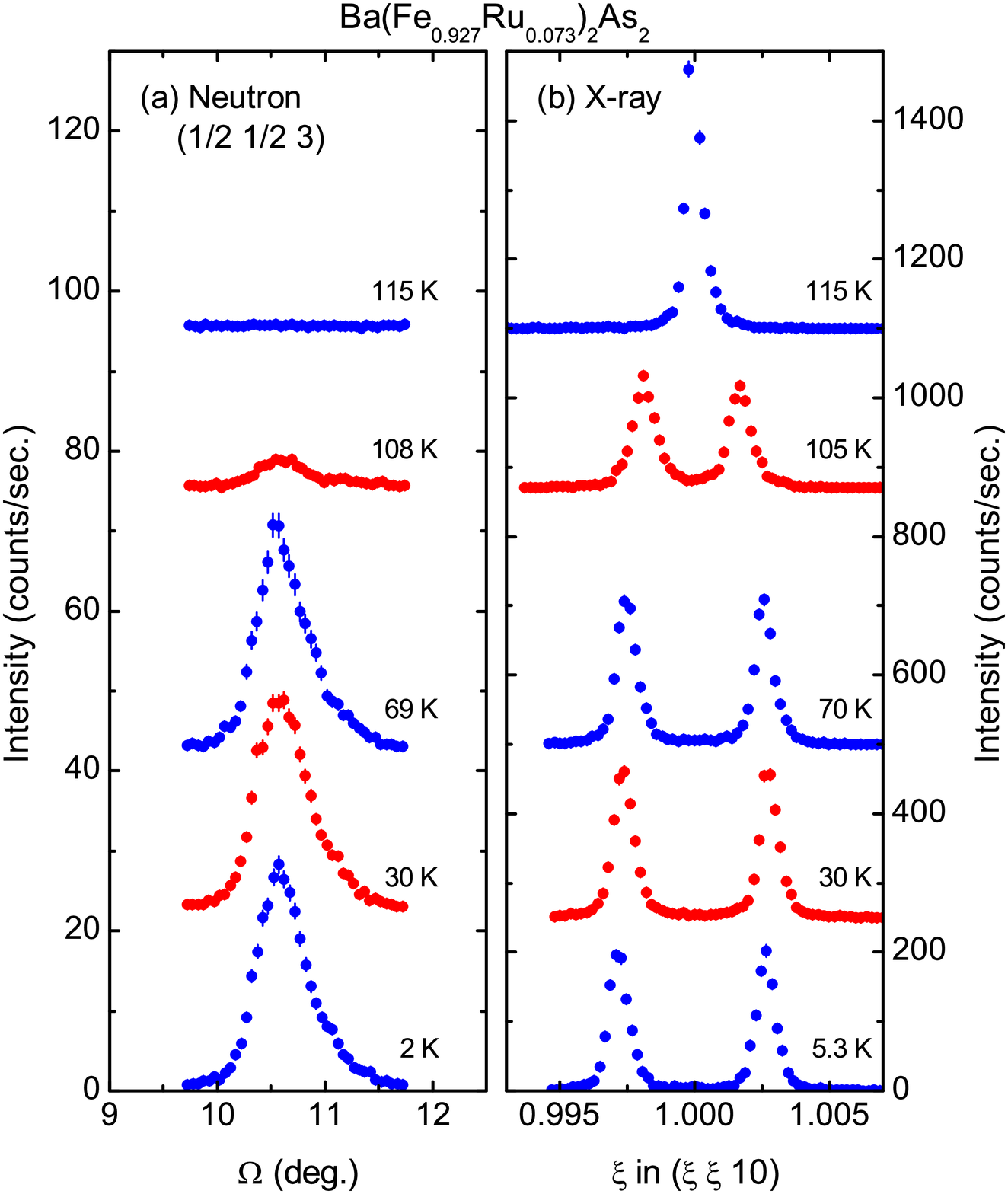}\\
\caption{(Color online) Temperature evolution of (a) the neutron
diffraction rocking scans through the ($\frac{1}{2}$ $\frac{1}{2}$
3) magnetic Bragg peak and (b) high-resolution x-ray diffraction
[$\xi \xi 0$]-scans through the (1 1 10) Bragg peak in
Ba(Fe$_{0.927}$Ru$_{0.073}$)$_2$As$_2$. For this sample $T_S$ =
$T_N$ = 109$\pm$1 K. The data are shown with arbitrary offsets.}
\label{scans}
\end{center}
\end{figure}

\begin{figure*}[t!]
\begin{center}
\includegraphics[clip, width=1.0\linewidth]{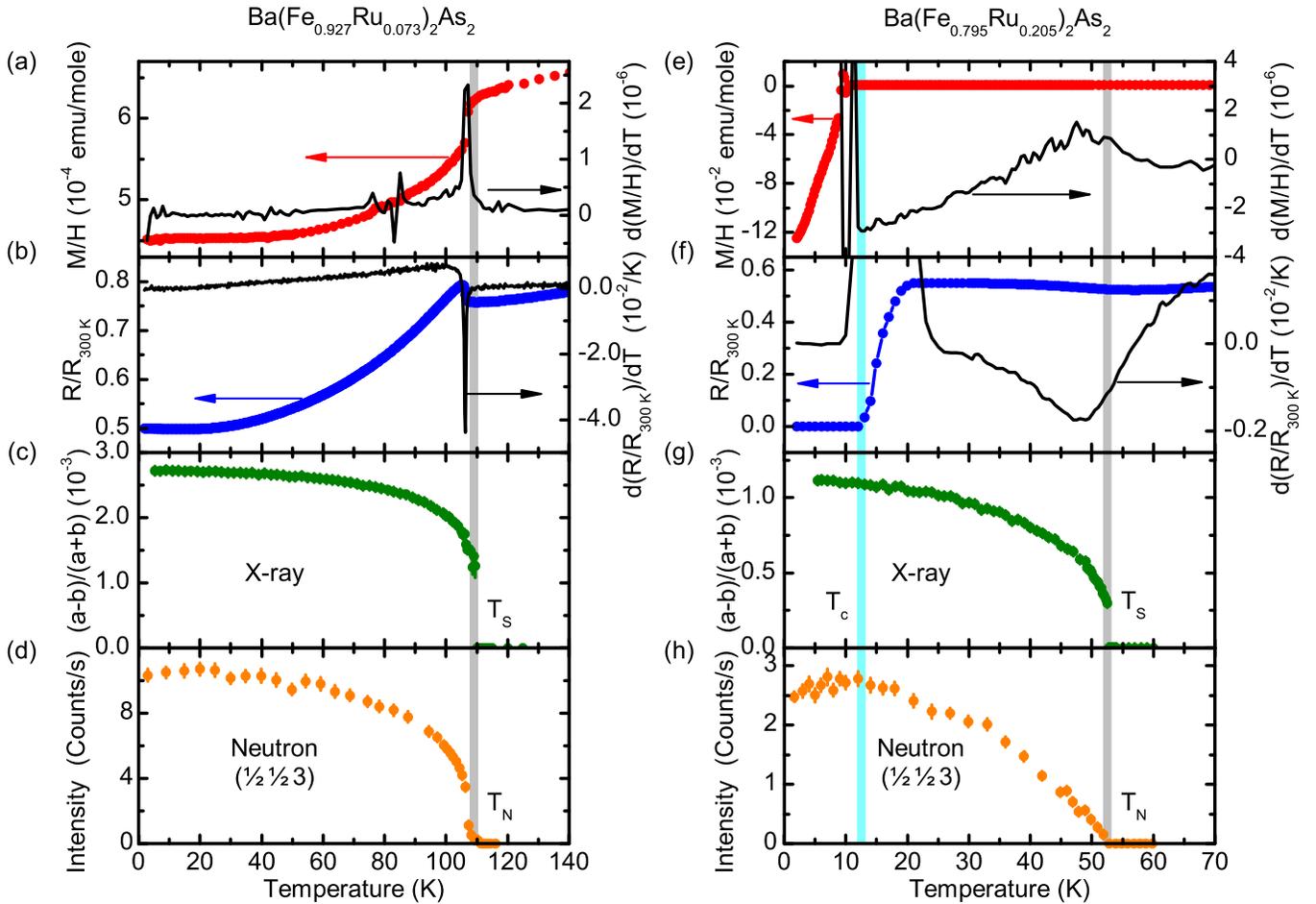}\\
\caption{(Color online) Plots of magnetization ($\frac{M}{H}$) and
its temperature derivative, $\frac{d(\frac{M}{H})}{dT}$, the
resistance ratio ($\frac{R}{R_{300~\textrm{K}}}$) and its
temperature derivative, the measured orthorhombic distortion
($\delta = \frac{a-b}{a+b}$), and the integrated magnetic intensity
at ($\frac{1}{2}$ $\frac{1}{2}$ 3) for
Ba(Fe$_{0.927}$Ru$_{0.073}$)$_2$As$_2$ in panels (a)-(d) and
Ba(Fe$_{0.795}$Ru$_{0.205}$)$_2$As$_2$ in panels (e)-(h).  For $x$ =
0.073 the measured magnetization, resistance and their derivatives
show sharp signatures at $T_S$ = $T_N$ = 107 K, close to the value
(109$\pm$1 K) measured by the x-ray and neutron scattering
measurements. For $x$ = 0.205, the signatures at $T_S$ = $T_N$ are
significantly broader. The maxima of the derivatives of the
magnetization and resistance are found at 49 K whereas the x-ray and
neutron scattering value is 52$\pm$1 K. } \label{all}
\end{center}
\end{figure*}

Neutron diffraction measurements were performed on the HB1A
diffractometer at the High Flux Isotope Reactor at Oak Ridge
National Laboratory using samples with a typical mass of
approximately 25 mg. The beam collimators before the
monochromator-between the monochromator and sample-between the
sample and analyzer-between the analyzer and detector were
48'-40'-40'-136'. HB1A operates at a fixed incident neutron energy
of 14.7 meV, and two pyrolytic graphite filters were employed to
effectively eliminate higher harmonics in the incident beam. The
samples were aligned such that the ($HHL$) reciprocal lattice plane
was coincident with the scattering plane of the spectrometer, and
were mounted in a closed-cycle refrigerator. The temperature
dependence of the scattering was studied at several nuclear Bragg
peak positions and at \textbf{Q}$_{\textrm{AFM}}$ = ($\frac{1}{2}$
$\frac{1}{2}$ $L$=odd) positions corresponding to the AFM order in
the parent and electron-doped BaFe$_2$As$_2$ compounds.

The high-resolution, single-crystal x-ray diffraction measurements
were performed on a four-circle diffractometer using Cu $K\alpha_1$
radiation from a rotating anode x-ray source, selected by a
germanium (111) monochromator. For the temperature-dependence
measurements, in addition to the parent BaFe$_2$As$_2$, we employed
the same single crystals of Ba(Fe$_{1-x}$Ru$_x$)$_2$As$_2$ ($x$ =
0.073 and 0.205) studied in our neutron measurements. The samples
were attached to a flat copper sample holder on the cold finger of a
closed-cycle displex refrigerator. The sample mosaicities were less
than 0.02\textsuperscript{$\circ$} full-width-at-half-maximum (FWHM)
as measured by rocking scans through the (1 1 10) reflection at room
temperature. The diffraction data were obtained as a function of
temperature between room temperature and 6 K, the base temperature
of the refrigerator.

\section{Results}

Figures~\ref{scans} (a) and (b) show neutron and x-ray data at
selected temperatures for Ba(Fe$_{1-x}$Ru$_x$)$_2$As$_2$ with $x$ =
0.073. Above $T_S$ = $T_N$ = 109$\pm$1 K, no scattering is observed
at Q$_{\textrm{AFM}}$ = ($\frac{1}{2}$ $\frac{1}{2}$ 3), but as the
temperature is lowered below $T_N$, the scattering increases
smoothly. The magnetic wave vector is identical to that for
BaFe$_2$As$_2$ compounds indicating that the magnetic structure is
the same AFM stripe-like structure observed for all AFM ordered
\emph{AE}Fe$_2$As$_2$ compounds (\emph{AE} = Ba, Sr, Ca), with AFM
alignment of the moments along the orthorhombic $\textbf{a}$ and
$\textbf{c}$ axes and FM alignment along the $\textbf{b}$ axis.
Analysis of the intensity ratios of different AFM reflections at
selected temperatures confirmed that the moment direction is along
the elongated orthorhombic $\textbf{a}$ direction. From our
high-resolution x-ray measurements we see [Fig.~\ref{scans} (b)]
that the (1 1 10) Bragg peak exhibits a sharp single peak above
$T_S$ = $T_N$ = 109$\pm$1 K consistent with a tetragonal structure
and splits into two peaks below $T_S$, characteristic of the
expected tetragonal-to-orthorhombic transition.

Figures~\ref{all} (a) and (b) summarize the magnetization and
resistance measurements on Ba(Fe$_{1-x}$Ru$_x$)$_2$As$_2$ with $x$ =
0.073. A sharp feature attributed to $T_S$/$T_N$ is observed at 107
K in the derivatives of magnetization and resistance. In
Fig.~\ref{all} (c) and (d), the orthorhombic distortion, $\delta =
\frac{a-b}{a+b}$, and the integrated magnetic scattering intensity,
measured from rocking scans through Q$_{\textrm{AFM}}$ =
($\frac{1}{2}$ $\frac{1}{2}$ 3), are plotted as a function of
temperature for $x$ = 0.073.  From these measurements we find that
$T_S$ = $T_N$ = 109$\pm$1 K, in reasonable agreement with the
thermodynamic and transport measurements given the inherent
uncertainty in assigning transition temperatures to features in the
magnetization and resistance. Figures~\ref{all} (e) and (f)
summarize the magnetization and resistance measurements on
Ba(Fe$_{1-x}$Ru$_x$)$_2$As$_2$ with $x$ = 0.205. Here, we see that
the characteristic features are much broader. According to the
criteria of Ref.~\onlinecite{thaler_physical_2010}, $T_S$/$T_N$ is
assigned to the maxima of the derivatives of magnetization and
resistance, which is 49 K. The x-ray and neutron data of
Figs.~\ref{all} (g) and (h) display the orthorhombic distortion
$\delta$ and the magnetic integrated intensity at Q$_{\textrm{AFM}}$
= ($\frac{1}{2}$ $\frac{1}{2}$ 3) for $x$ = 0.205 and yield $T_S$ =
$T_N$ = 52$\pm$1 K. The transition temperatures derived from the
criteria of Ref.~\onlinecite{thaler_physical_2010} are up to 3 K
lower than the observed transition temperatures derived from the
x-ray and neutron diffraction measurements. Most importantly,
however, we find that, within experimental error, the structural and
magnetic transitions remain locked together with increasing Ru
doping and this behavior clearly differs from that found for the
electron-doped compounds.

\begin{figure}
\begin{center}
\includegraphics[clip, width=1.0\linewidth]{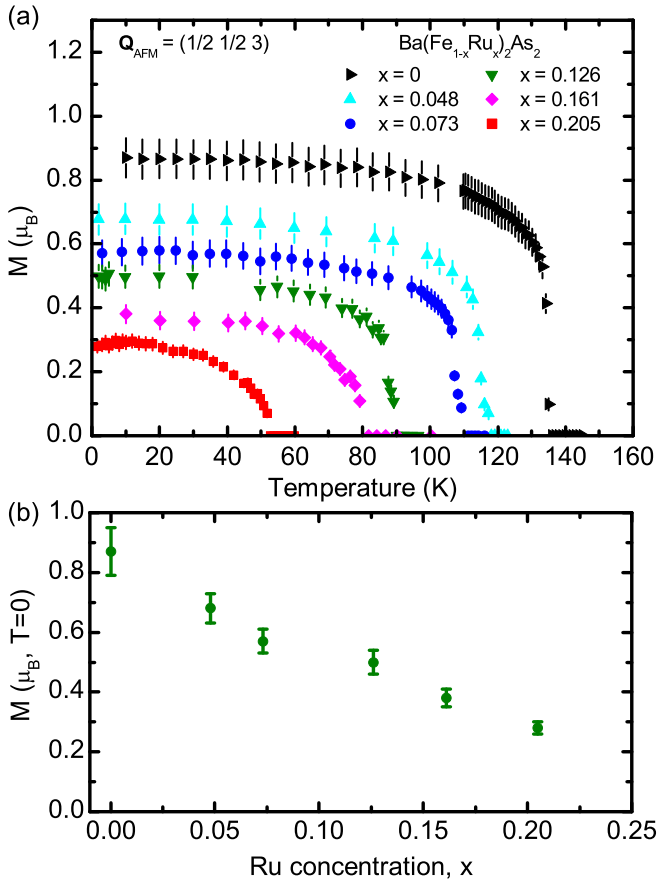}\\
\caption{(Color online) (a) Temperature dependence of the ordered
magnetic moment calculated from the integrated intensity of the
($\frac{1}{2}$ $\frac{1}{2}$ 3) magnetic Bragg peak from
Ba(Fe$_{1-x}$Ru$_{x}$)$_2$As$_2$. (b) The extrapolated ordered
moment at zero temperature as a function of Ru concentration, $x$. }
\label{moment}
\end{center}
\end{figure}

\begin{figure}
\begin{center}
\includegraphics[clip, width=1.0\linewidth]{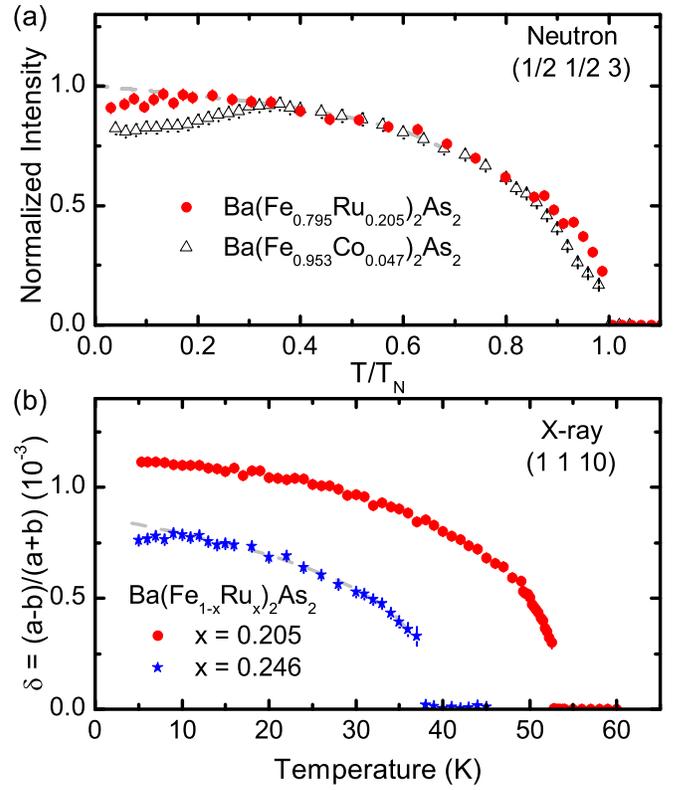}\\
\caption{(Color online)(a) Comparison of the suppression of AFM
order below $T_c$ between the 20.5\% Ru (filled circles) and the
4.7\% Co (open triangles)~\cite{fernandes_unconventional_2010} doped
BaFe$_2$As$_2$ samples. Intensities are normalized for comparison.
(b) Orthorhombic distortion for Ba(Fe$_{1-x}$Ru$_x$)$_2$As$_2$ with
$x$ = 0.205 (circles) and 0.246 (stars). The reduction in the
distortion below $T_c$ is not clearly observable for $x$ = 0.205 but
it is evident for $x$ = 0.246. The gray dashed lines are guides for
eyes. } \label{comp}
\end{center}
\end{figure}

Ba(Fe$_{1-x}$Ru$_x$)$_2$As$_2$ crystals with $x$ = 0, 0.048, 0.126,
and 0.161 were also examined by neutron diffraction and the results
for the entire series are summarized in Fig.~\ref{moment}. The
magnetic integrated intensities were, again, determined from rocking
scans through the magnetic peak at ($\frac{1}{2}$ $\frac{1}{2}$ 3)
as a function of temperature and put on an absolute basis using the
known mass of the samples and the magnetic diffraction from the
parent compound, BaFe$_2$As$_2$, measured under identical
conditions.\cite{fernandes_unconventional_2010} The ordered moment
as a function of temperature for each sample is presented in
Fig.~\ref{moment}(a), and the ordered moments extrapolated to $T =
0$ are shown in Fig.~\ref{moment}(b). We see that as the Ru
concentration increases, the ordered moment decreases monotonically.

Turning now to the effects of superconductivity on the AFM ordering
and structural distortion, we first note that for the $x$ = 0.205
sample, the resistance and magnetization data show the existence of
superconductivity below $T_c$ $\approx$ 13 K in Figs.~\ref{all} (e)
and (f). For this sample, in Fig.~2 (h), we observe a suppression of
the AFM order below $T_c$ similar to what has been reported
previously for Co-, Rh-, and Ni-doped
BaFe$_2$As$_2$,\cite{pratt_coexistence_2009,kreyssig_suppression_2010,wang_electron-doping_2010}
where the presence of both AFM and superconductivity has been
attributed to microscopically coexisting states that compete for the
same itinerant electrons. It has also been established that the
onset of superconductivity leads to a suppression of the
orthorhombic distortion in the electron-doped compounds.
Refs.~\onlinecite{nandi_anomalous_2010} and
~\onlinecite{kreyssig_suppression_2010}, for example, described this
effect below $T_c$ for both Co- and Rh-doped BaFe$_2$As$_2$,
respectively. Because $\frac{T_c}{T_N}$ for
Ba(Fe$_{0.795}$Ru$_{0.205}$)$_2$As$_2$ is approximately half the
value of $\frac{T_c}{T_N}$ for
Ba(Fe$_{0.953}$Co$_{0.047}$)$_2$As$_2$, the magnitude of suppression
of AFM order at the base temperature of our measurement is
correspondingly smaller [Fig.~\ref{comp} (a)], and the reduction of
the orthorhombic distortion is not clearly observed [Figs.~\ref{all}
(g) and ~\ref{comp}(b)]. Therefore, we have also studied an
additional concentration, $x$ = 0.246$\pm$0.005 ($T_c$ $\approx$ 14
K), by high-resolution x-ray diffraction and, as shown in
Fig.~\ref{comp} (b), see the suppression of the orthorhombic
distortion below $T_c$.

\section{Discussion and summary}

\begin{figure}[!ht]
\begin{center}
\includegraphics[clip, width=1.0\linewidth]{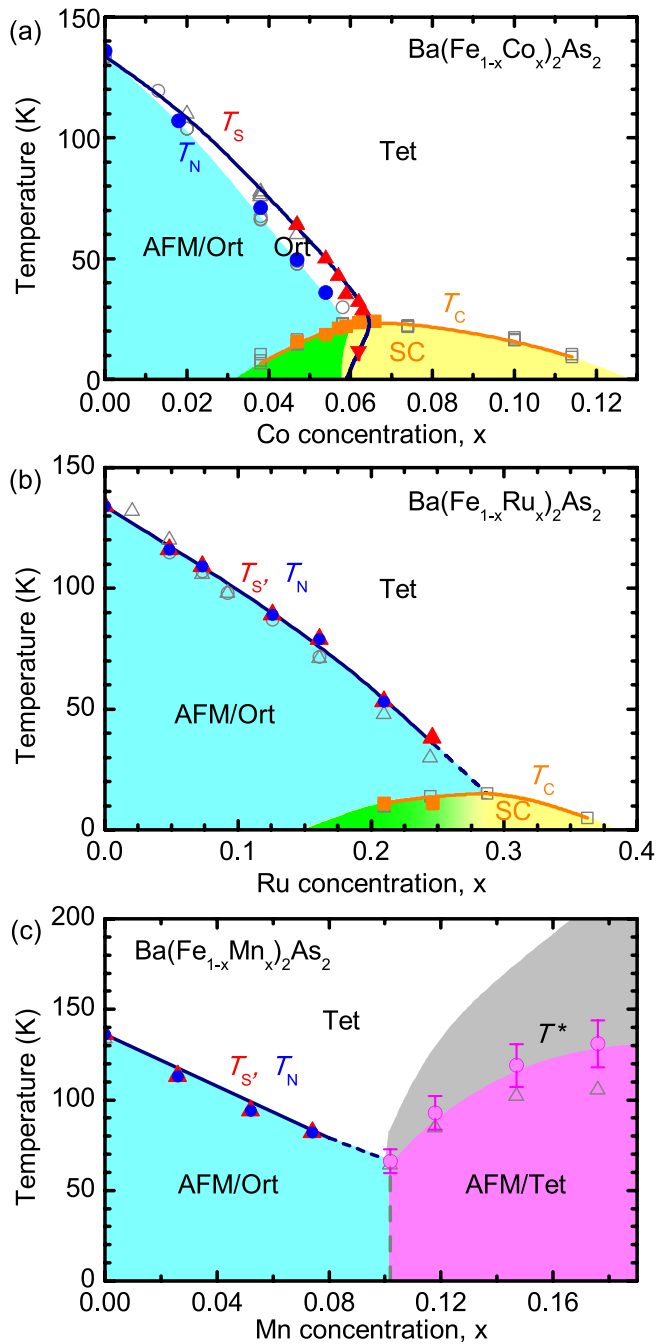}\\
\caption{(Color online) Compositional phase diagrams for (a)
Ba(Fe$_{1-x}$Co$_x$)$_2$As$_2$ from
Ref.~\onlinecite{nandi_anomalous_2010}, (b)
Ba(Fe$_{1-x}$Ru$_x$)$_2$As$_2$ from the present work and
Ref.~\onlinecite{thaler_physical_2010}, and (c)
Ba(Fe$_{1-x}$Mn$_x$)$_2$As$_2$ from Ref.~\onlinecite{mgkim_mn_2010}.
The gray open triangles and open circles denote data taken from
resistance and magnetization data respectively.  The gray open
squares denote bulk measurements of $T_c$.  Filled red triangles
denote $T_S$ measured by x-ray diffraction, filled blue circles
denote $T_N$ measured by neutron diffraction, and the filled orange
squares represent values for $T_c$ from the x-ray and neutron data.
 Filled magneta circels denote $T^*$ determined for the Mn doped sample by neutron measurements (see Ref.~\onlinecite{mgkim_mn_2010}).}\label{PD}
\end{center}
\end{figure}

Together with our previous investigations of
Ba(Fe$_{1-x}$Co$_x$)$_2$As$_2$ and Ba(Fe$_{1-x}$Mn$_x$)$_2$As$_2$,
we now have a more complete picture of the effects of electron, hole
and isoelectronic doping on the Fe site in the BaFe$_2$As$_2$
compound. The compositional phase diagrams for all three doping
series are shown in Fig.~\ref{PD}. Summarizing the trends
illustrated in Fig.~\ref{PD} (a) we see that for the Co-doped
series, at low doping, the magnetic and structural transitions split
with increasing Co concentration, superconductivity emerges over a
finite compositional range and coexists with AFM order over an even
more limited range of Co doping.  The back-bending of the AFM and
structural distortion phase lines in the superconducting region
identify the reentrance of the paramagnetic and tetragonal phases at
low temperature.  Figs.~\ref{PD} (a) and (b) display both the
similarities and differences between Ba(Fe$_{1-x}$Co$_x$)$_2$As$_2$
and Ba(Fe$_{1-x}$Ru$_x$)$_2$As$_2$. As found for Co substitution, Ru
doping results in the suppression of the AFM and structural
transitions and superconductivity emerges over a finite range of Ru
concentration.  The suppression of both the AFM order and
orthorhombic distortion in the superconducting region suggests that
reentrance of the paramagnetic tetragonal phase may also be found at
some Ru doping concentration as well. However, for Ru doping the AFM
and structural transitions remain locked together over an extended
compositional range with respect to the phase diagram for Co doping.
In Fig.~\ref{PD} (c), we reproduce the compositional phase diagram
for Mn doping, which is quite different from what is found for
either Co or Ru substitution on the Fe site. Superconductivity is
not in evidence at any Mn concentration and, while the AFM and
structural transitions remain locked together with increasing Mn
concentration, as found for Ba(Fe$_{1-x}$Ru$_x$)$_2$As$_2$, the
structural distortion abruptly disappears for Mn doping in excess of
$x > 0.102$ although the AFM Bragg peak characteristic of
stripe-like ordering persists. The latter observation is quite
puzzling since all models for stripe-like ordering in the iron
arsenides anticipate an attendant orthorhombic distortion due to
magnetoelastic effects.  However, we have previously proposed that
the scattering at \textbf{Q}$_{\textrm{AFM}}$ = ($\frac{1}{2}$
$\frac{1}{2}$ $L$=odd) positions may also be explained by the
presence of a two-\textbf{Q} magnetic structure that is again
consistent with tetragonal symmetry.\cite{mgkim_mn_2010}

It is clear that the interactions associated with structural,
magnetic and superconducting instabilities in the
\emph{AE}Fe$_2$As$_2$ compounds are finely balanced and can be
readily tuned through chemical substitution as well as pressure. For
example, similarities between chemical doping and pressure were
previously discussed for K doping on the Ba
site.\cite{kimber_similarities_2009} For electron doping on the Fe
site, a rigid band picture appears to be applicable, at least to
first order, in explaining the phenomenology of magnetism, structure
and superconductivity.  Doping with Mn, however, clearly introduces
strong perturbations on both the electronic and chemical structure,
likely as a consequence of the more localized nature of the Mn
magnetic moment.  Doping with Ru provides a new interesting case
study where, nominally, no electrons or holes are added to the
system although the first band-structure calculations indicated that
Ru substitution introduces additional electron
carriers.\cite{sharma_superconductivity_2010} However, Hall effect
and angle-resolved photoemission spectroscopy
measurements\cite{rullier-albenque_hole_2010,Brouet_2010}, have
shown that the Ru substitution does not induce electron or hole
doping, but does strongly modify the electronic structure by
increasing both the number of carriers and their mobility by
reducing correlation effects.

Summarizing, we have presented a systematic investigation of the AFM
ordering and structural distortion on the series of
Ba(Fe$_{1-x}$Ru$_x$)$_2$As$_2$ compounds ($0 \leq x \leq 0.246$).
Our neutron and x-ray diffraction measurements demonstrate that,
unlike the behavior found for the electron-doped compounds, the
structural and magnetic transitions remain coincident in
temperature, as also observed for low Mn doping.  Both the magnetic
and structural transitions are gradually suppressed with increased
Ru concentration but, in contrast to the case for Mn doping where
superconductivity is absent, AFM order coexists with
superconductivity. In the superconducting samples, we find a strong
competition between superconductivity, the AFM ordering and the
structural distortion.

We acknowledge valuable discussions with J. Schmalian and R. M.
Fernandes. This work was supported by the Division of Materials
Sciences and Engineering, Office of Basic Energy Sciences, U.S.
Department of Energy. Ames Laboratory is operated for the U.S.
Department of Energy by Iowa State University under Contract No.
DE-AC02-07CH11358. The work at the High Flux Isotope Reactor, Oak
Ridge National Laboratory (ORNL), was sponsored by the Scientific
User Facilities Division, Office of Basic Energy Sciences, U.S.
Department of Energy (U.S. DOE). ORNL is operated by UT-Battelle,
LLC for the U.S. DOE under Contract No. DE-AC05-00OR22725.

\bibliographystyle{apsrev}
\bibliography{elastic_ru_aig}

\begin{thebibliography}{42}
\expandafter\ifx\csname natexlab\endcsname\relax\def\natexlab#1{#1}\fi
\expandafter\ifx\csname bibnamefont\endcsname\relax
  \def\bibnamefont#1{#1}\fi
\expandafter\ifx\csname bibfnamefont\endcsname\relax
  \def\bibfnamefont#1{#1}\fi
\expandafter\ifx\csname citenamefont\endcsname\relax
  \def\citenamefont#1{#1}\fi
\expandafter\ifx\csname url\endcsname\relax
  \def\url#1{\texttt{#1}}\fi
\expandafter\ifx\csname urlprefix\endcsname\relax\def\urlprefix{URL }\fi
\providecommand{\bibinfo}[2]{#2}
\providecommand{\eprint}[2][]{\url{#2}}

\bibitem[{\citenamefont{Kamihara et~al.}(2008)\citenamefont{Kamihara, Watanabe,
  Hirano, and Hosono}}]{kamihara_iron-based_2008}
\bibinfo{author}{\bibfnamefont{Y.}~\bibnamefont{Kamihara}},
  \bibinfo{author}{\bibfnamefont{T.}~\bibnamefont{Watanabe}},
  \bibinfo{author}{\bibfnamefont{M.}~\bibnamefont{Hirano}}, \bibnamefont{and}
  \bibinfo{author}{\bibfnamefont{H.}~\bibnamefont{Hosono}},
  \bibinfo{journal}{J. Am. Chem. Soc.} \textbf{\bibinfo{volume}{130}},
  \bibinfo{pages}{3296} (\bibinfo{year}{2008}).

\bibitem[{\citenamefont{Rotter et~al.}(2008{\natexlab{a}})\citenamefont{Rotter,
  Tegel, and Johrendt}}]{rotter_superconductivity_2008}
\bibinfo{author}{\bibfnamefont{M.}~\bibnamefont{Rotter}},
  \bibinfo{author}{\bibfnamefont{M.}~\bibnamefont{Tegel}}, \bibnamefont{and}
  \bibinfo{author}{\bibfnamefont{D.}~\bibnamefont{Johrendt}},
  \bibinfo{journal}{Phys. Rev. Lett.} \textbf{\bibinfo{volume}{101}},
  \bibinfo{pages}{107006} (\bibinfo{year}{2008}{\natexlab{a}}).

\bibitem[{\citenamefont{Rotter et~al.}(2008{\natexlab{b}})\citenamefont{Rotter,
  Tegel, Johrendt, Schellenberg, Hermes, and
  P\"{o}ttgen}}]{rotter_spin-density-wave_2008}
\bibinfo{author}{\bibfnamefont{M.}~\bibnamefont{Rotter}},
  \bibinfo{author}{\bibfnamefont{M.}~\bibnamefont{Tegel}},
  \bibinfo{author}{\bibfnamefont{D.}~\bibnamefont{Johrendt}},
  \bibinfo{author}{\bibfnamefont{I.}~\bibnamefont{Schellenberg}},
  \bibinfo{author}{\bibfnamefont{W.}~\bibnamefont{Hermes}}, \bibnamefont{and}
  \bibinfo{author}{\bibfnamefont{R.}~\bibnamefont{P\"{o}ttgen}},
  \bibinfo{journal}{Phys. Rev. B} \textbf{\bibinfo{volume}{78}},
  \bibinfo{pages}{020503(R)} (\bibinfo{year}{2008}{\natexlab{b}}).

\bibitem[{\citenamefont{Huang et~al.}(2008)\citenamefont{Huang, Qiu, Bao,
  Green, Lynn, Gasparovic, Wu, Wu, and Chen}}]{huang_neutron-diffraction_2008}
\bibinfo{author}{\bibfnamefont{Q.}~\bibnamefont{Huang}},
  \bibinfo{author}{\bibfnamefont{Y.}~\bibnamefont{Qiu}},
  \bibinfo{author}{\bibfnamefont{W.}~\bibnamefont{Bao}},
  \bibinfo{author}{\bibfnamefont{M.~A.} \bibnamefont{Green}},
  \bibinfo{author}{\bibfnamefont{J.~W.} \bibnamefont{Lynn}},
  \bibinfo{author}{\bibfnamefont{Y.~C.} \bibnamefont{Gasparovic}},
  \bibinfo{author}{\bibfnamefont{T.}~\bibnamefont{Wu}},
  \bibinfo{author}{\bibfnamefont{G.}~\bibnamefont{Wu}}, \bibnamefont{and}
  \bibinfo{author}{\bibfnamefont{X.~H.} \bibnamefont{Chen}},
  \bibinfo{journal}{Phys. Rev. Lett.} \textbf{\bibinfo{volume}{101}},
  \bibinfo{pages}{257003} (\bibinfo{year}{2008}).

\bibitem[{\citenamefont{Jesche et~al.}(2008)\citenamefont{Jesche,
  {Caroca-Canales}, Rosner, Borrmann, Ormeci, Kasinathan, Klauss, Luetkens,
  Khasanov, Amato et~al.}}]{jesche_strong_2008}
\bibinfo{author}{\bibfnamefont{A.}~\bibnamefont{Jesche}},
  \bibinfo{author}{\bibfnamefont{N.}~\bibnamefont{{Caroca-Canales}}},
  \bibinfo{author}{\bibfnamefont{H.}~\bibnamefont{Rosner}},
  \bibinfo{author}{\bibfnamefont{H.}~\bibnamefont{Borrmann}},
  \bibinfo{author}{\bibfnamefont{A.}~\bibnamefont{Ormeci}},
  \bibinfo{author}{\bibfnamefont{D.}~\bibnamefont{Kasinathan}},
  \bibinfo{author}{\bibfnamefont{H.~H.} \bibnamefont{Klauss}},
  \bibinfo{author}{\bibfnamefont{H.}~\bibnamefont{Luetkens}},
  \bibinfo{author}{\bibfnamefont{R.}~\bibnamefont{Khasanov}},
  \bibinfo{author}{\bibfnamefont{A.}~\bibnamefont{Amato}},
  \bibnamefont{et~al.}, \bibinfo{journal}{Phys. Rev. B}
  \textbf{\bibinfo{volume}{78}}, \bibinfo{pages}{180504(R)}
  (\bibinfo{year}{2008}).

\bibitem[{\citenamefont{Goldman et~al.}(2008)\citenamefont{Goldman, Argyriou,
  Ouladdiaf, Chatterji, Kreyssig, Nandi, Ni, Bud'ko, Canfield, and
  {McQueeney}}}]{goldman_lattice_2008}
\bibinfo{author}{\bibfnamefont{A.~I.} \bibnamefont{Goldman}},
  \bibinfo{author}{\bibfnamefont{D.~N.} \bibnamefont{Argyriou}},
  \bibinfo{author}{\bibfnamefont{B.}~\bibnamefont{Ouladdiaf}},
  \bibinfo{author}{\bibfnamefont{T.}~\bibnamefont{Chatterji}},
  \bibinfo{author}{\bibfnamefont{A.}~\bibnamefont{Kreyssig}},
  \bibinfo{author}{\bibfnamefont{S.}~\bibnamefont{Nandi}},
  \bibinfo{author}{\bibfnamefont{N.}~\bibnamefont{Ni}},
  \bibinfo{author}{\bibfnamefont{S.~L.} \bibnamefont{Bud'ko}},
  \bibinfo{author}{\bibfnamefont{P.~C.} \bibnamefont{Canfield}},
  \bibnamefont{and} \bibinfo{author}{\bibfnamefont{R.~J.}
  \bibnamefont{{McQueeney}}}, \bibinfo{journal}{Phys. Rev. B}
  \textbf{\bibinfo{volume}{78}}, \bibinfo{pages}{100506(R)}
  (\bibinfo{year}{2008}).

\bibitem[{\citenamefont{Chen et~al.}(2009)\citenamefont{Chen, Ren, Qiu, Bao,
  Liu, Wu, Wu, Xie, Wang, Huang et~al.}}]{chen_coexistence_2009}
\bibinfo{author}{\bibfnamefont{H.}~\bibnamefont{Chen}},
  \bibinfo{author}{\bibfnamefont{Y.}~\bibnamefont{Ren}},
  \bibinfo{author}{\bibfnamefont{Y.}~\bibnamefont{Qiu}},
  \bibinfo{author}{\bibfnamefont{W.}~\bibnamefont{Bao}},
  \bibinfo{author}{\bibfnamefont{R.~H.} \bibnamefont{Liu}},
  \bibinfo{author}{\bibfnamefont{G.}~\bibnamefont{Wu}},
  \bibinfo{author}{\bibfnamefont{T.}~\bibnamefont{Wu}},
  \bibinfo{author}{\bibfnamefont{Y.~L.} \bibnamefont{Xie}},
  \bibinfo{author}{\bibfnamefont{X.~F.} \bibnamefont{Wang}},
  \bibinfo{author}{\bibfnamefont{Q.}~\bibnamefont{Huang}},
  \bibnamefont{et~al.}, \bibinfo{journal}{Europhys. Lett.}
  \textbf{\bibinfo{volume}{85}}, \bibinfo{pages}{17006} (\bibinfo{year}{2009}).

\bibitem[{\citenamefont{Inosov et~al.}(2009)\citenamefont{Inosov, Leineweber,
  Yang, Park, Christensen, Dinnebier, Sun, Niedermayer, Haug, Stephens
  et~al.}}]{inosov_suppression_2009}
\bibinfo{author}{\bibfnamefont{D.~S.} \bibnamefont{Inosov}},
  \bibinfo{author}{\bibfnamefont{A.}~\bibnamefont{Leineweber}},
  \bibinfo{author}{\bibfnamefont{X.}~\bibnamefont{Yang}},
  \bibinfo{author}{\bibfnamefont{J.~T.} \bibnamefont{Park}},
  \bibinfo{author}{\bibfnamefont{N.~B.} \bibnamefont{Christensen}},
  \bibinfo{author}{\bibfnamefont{R.}~\bibnamefont{Dinnebier}},
  \bibinfo{author}{\bibfnamefont{G.~L.} \bibnamefont{Sun}},
  \bibinfo{author}{\bibfnamefont{C.}~\bibnamefont{Niedermayer}},
  \bibinfo{author}{\bibfnamefont{D.}~\bibnamefont{Haug}},
  \bibinfo{author}{\bibfnamefont{P.~W.} \bibnamefont{Stephens}},
  \bibnamefont{et~al.}, \bibinfo{journal}{Phys. Rev. B}
  \textbf{\bibinfo{volume}{79}}, \bibinfo{pages}{224503}
  (\bibinfo{year}{2009}).

\bibitem[{\citenamefont{Saha et~al.}(2010)\citenamefont{Saha, Drye,
  Kirshenbaum, Butch, Zavalij, and Paglione}}]{saha_superconductivity_2010}
\bibinfo{author}{\bibfnamefont{S.~R.} \bibnamefont{Saha}},
  \bibinfo{author}{\bibfnamefont{T.}~\bibnamefont{Drye}},
  \bibinfo{author}{\bibfnamefont{K.}~\bibnamefont{Kirshenbaum}},
  \bibinfo{author}{\bibfnamefont{N.~P.} \bibnamefont{Butch}},
  \bibinfo{author}{\bibfnamefont{P.~Y.} \bibnamefont{Zavalij}},
  \bibnamefont{and} \bibinfo{author}{\bibfnamefont{J.}~\bibnamefont{Paglione}},
  \bibinfo{journal}{J. Phys.: Condens. Matter} \textbf{\bibinfo{volume}{22}},
  \bibinfo{pages}{072204} (\bibinfo{year}{2010}).

\bibitem[{\citenamefont{Sefat et~al.}(2008)\citenamefont{Sefat, Jin, {McGuire},
  Sales, Singh, and Mandrus}}]{sefat_superconductivity_2008}
\bibinfo{author}{\bibfnamefont{A.~S.} \bibnamefont{Sefat}},
  \bibinfo{author}{\bibfnamefont{R.}~\bibnamefont{Jin}},
  \bibinfo{author}{\bibfnamefont{M.~A.} \bibnamefont{{McGuire}}},
  \bibinfo{author}{\bibfnamefont{B.~C.} \bibnamefont{Sales}},
  \bibinfo{author}{\bibfnamefont{D.~J.} \bibnamefont{Singh}}, \bibnamefont{and}
  \bibinfo{author}{\bibfnamefont{D.}~\bibnamefont{Mandrus}},
  \bibinfo{journal}{Phys. Rev. Lett.} \textbf{\bibinfo{volume}{101}},
  \bibinfo{pages}{117004} (\bibinfo{year}{2008}).

\bibitem[{\citenamefont{Ni et~al.}(2008)\citenamefont{Ni, Tillman, Yan,
  Kracher, Hannahs, Bud'ko, and Canfield}}]{ni_effects_2008}
\bibinfo{author}{\bibfnamefont{N.}~\bibnamefont{Ni}},
  \bibinfo{author}{\bibfnamefont{M.~E.} \bibnamefont{Tillman}},
  \bibinfo{author}{\bibfnamefont{J.-Q.} \bibnamefont{Yan}},
  \bibinfo{author}{\bibfnamefont{A.}~\bibnamefont{Kracher}},
  \bibinfo{author}{\bibfnamefont{S.~T.} \bibnamefont{Hannahs}},
  \bibinfo{author}{\bibfnamefont{S.~L.} \bibnamefont{Bud'ko}},
  \bibnamefont{and} \bibinfo{author}{\bibfnamefont{P.~C.}
  \bibnamefont{Canfield}}, \bibinfo{journal}{Phys. Rev. B}
  \textbf{\bibinfo{volume}{78}}, \bibinfo{pages}{214515}
  (\bibinfo{year}{2008}).

\bibitem[{\citenamefont{Lester et~al.}(2009)\citenamefont{Lester, Chu,
  Analytis, Capelli, Erickson, Condron, Toney, Fisher, and
  Hayden}}]{lester_2009}
\bibinfo{author}{\bibfnamefont{C.}~\bibnamefont{Lester}},
  \bibinfo{author}{\bibfnamefont{J.-H.} \bibnamefont{Chu}},
  \bibinfo{author}{\bibfnamefont{J.~G.} \bibnamefont{Analytis}},
  \bibinfo{author}{\bibfnamefont{S.~C.} \bibnamefont{Capelli}},
  \bibinfo{author}{\bibfnamefont{A.~S.} \bibnamefont{Erickson}},
  \bibinfo{author}{\bibfnamefont{C.~L.} \bibnamefont{Condron}},
  \bibinfo{author}{\bibfnamefont{M.~F.} \bibnamefont{Toney}},
  \bibinfo{author}{\bibfnamefont{I.~R.} \bibnamefont{Fisher}},
  \bibnamefont{and} \bibinfo{author}{\bibfnamefont{S.~M.}
  \bibnamefont{Hayden}}, \bibinfo{journal}{Phys. Rev. B}
  \textbf{\bibinfo{volume}{79}}, \bibinfo{pages}{144523}
  (\bibinfo{year}{2009}).

\bibitem[{\citenamefont{Li et~al.}(2009)\citenamefont{Li, Luo, Wang, Chen, Ren,
  Tao, Li, Lin, He, Zhu et~al.}}]{li_superconductivity_2009}
\bibinfo{author}{\bibfnamefont{L.~J.} \bibnamefont{Li}},
  \bibinfo{author}{\bibfnamefont{Y.~K.} \bibnamefont{Luo}},
  \bibinfo{author}{\bibfnamefont{Q.~B.} \bibnamefont{Wang}},
  \bibinfo{author}{\bibfnamefont{H.}~\bibnamefont{Chen}},
  \bibinfo{author}{\bibfnamefont{Z.}~\bibnamefont{Ren}},
  \bibinfo{author}{\bibfnamefont{Q.}~\bibnamefont{Tao}},
  \bibinfo{author}{\bibfnamefont{Y.~K.} \bibnamefont{Li}},
  \bibinfo{author}{\bibfnamefont{X.}~\bibnamefont{Lin}},
  \bibinfo{author}{\bibfnamefont{M.}~\bibnamefont{He}},
  \bibinfo{author}{\bibfnamefont{Z.~W.} \bibnamefont{Zhu}},
  \bibnamefont{et~al.}, \bibinfo{journal}{New J. Phys.}
  \textbf{\bibinfo{volume}{11}}, \bibinfo{pages}{025008}
  (\bibinfo{year}{2009}).

\bibitem[{\citenamefont{Ni et~al.}(2009)\citenamefont{Ni, Thaler, Kracher, Yan,
  Bud'ko, and Canfield}}]{ni_phase_2009}
\bibinfo{author}{\bibfnamefont{N.}~\bibnamefont{Ni}},
  \bibinfo{author}{\bibfnamefont{A.}~\bibnamefont{Thaler}},
  \bibinfo{author}{\bibfnamefont{A.}~\bibnamefont{Kracher}},
  \bibinfo{author}{\bibfnamefont{J.~Q.} \bibnamefont{Yan}},
  \bibinfo{author}{\bibfnamefont{S.~L.} \bibnamefont{Bud'ko}},
  \bibnamefont{and} \bibinfo{author}{\bibfnamefont{P.~C.}
  \bibnamefont{Canfield}}, \bibinfo{journal}{Phys. Rev. B}
  \textbf{\bibinfo{volume}{80}}, \bibinfo{pages}{024511}
  (\bibinfo{year}{2009}).

\bibitem[{\citenamefont{Harriger et~al.}(2009)\citenamefont{Harriger,
  Schneidewind, Li, Zhao, Li, Lu, Dong, Zhou, Zhao, Hu
  et~al.}}]{harriger_transition_2009}
\bibinfo{author}{\bibfnamefont{L.~W.} \bibnamefont{Harriger}},
  \bibinfo{author}{\bibfnamefont{A.}~\bibnamefont{Schneidewind}},
  \bibinfo{author}{\bibfnamefont{S.}~\bibnamefont{Li}},
  \bibinfo{author}{\bibfnamefont{J.}~\bibnamefont{Zhao}},
  \bibinfo{author}{\bibfnamefont{Z.}~\bibnamefont{Li}},
  \bibinfo{author}{\bibfnamefont{W.}~\bibnamefont{Lu}},
  \bibinfo{author}{\bibfnamefont{X.}~\bibnamefont{Dong}},
  \bibinfo{author}{\bibfnamefont{F.}~\bibnamefont{Zhou}},
  \bibinfo{author}{\bibfnamefont{Z.}~\bibnamefont{Zhao}},
  \bibinfo{author}{\bibfnamefont{J.}~\bibnamefont{Hu}}, \bibnamefont{et~al.},
  \bibinfo{journal}{Phys. Rev. Lett.} \textbf{\bibinfo{volume}{103}},
  \bibinfo{pages}{087005} (\bibinfo{year}{2009}).

\bibitem[{\citenamefont{Pratt et~al.}(2009)\citenamefont{Pratt, Tian, Kreyssig,
  Zarestky, Nandi, Ni, Bud'ko, Canfield, Goldman, and
  {McQueeney}}}]{pratt_coexistence_2009}
\bibinfo{author}{\bibfnamefont{D.~K.} \bibnamefont{Pratt}},
  \bibinfo{author}{\bibfnamefont{W.}~\bibnamefont{Tian}},
  \bibinfo{author}{\bibfnamefont{A.}~\bibnamefont{Kreyssig}},
  \bibinfo{author}{\bibfnamefont{J.~L.} \bibnamefont{Zarestky}},
  \bibinfo{author}{\bibfnamefont{S.}~\bibnamefont{Nandi}},
  \bibinfo{author}{\bibfnamefont{N.}~\bibnamefont{Ni}},
  \bibinfo{author}{\bibfnamefont{S.~L.} \bibnamefont{Bud'ko}},
  \bibinfo{author}{\bibfnamefont{P.~C.} \bibnamefont{Canfield}},
  \bibinfo{author}{\bibfnamefont{A.~I.} \bibnamefont{Goldman}},
  \bibnamefont{and} \bibinfo{author}{\bibfnamefont{R.~J.}
  \bibnamefont{{McQueeney}}}, \bibinfo{journal}{Phys. Rev. Lett.}
  \textbf{\bibinfo{volume}{103}}, \bibinfo{pages}{087001}
  (\bibinfo{year}{2009}).

\bibitem[{\citenamefont{Kreyssig et~al.}(2010)\citenamefont{Kreyssig, Kim,
  Nandi, Pratt, Tian, Zarestky, Ni, Thaler, Bud'ko, Canfield
  et~al.}}]{kreyssig_suppression_2010}
\bibinfo{author}{\bibfnamefont{A.}~\bibnamefont{Kreyssig}},
  \bibinfo{author}{\bibfnamefont{M.~G.} \bibnamefont{Kim}},
  \bibinfo{author}{\bibfnamefont{S.}~\bibnamefont{Nandi}},
  \bibinfo{author}{\bibfnamefont{D.~K.} \bibnamefont{Pratt}},
  \bibinfo{author}{\bibfnamefont{W.}~\bibnamefont{Tian}},
  \bibinfo{author}{\bibfnamefont{J.~L.} \bibnamefont{Zarestky}},
  \bibinfo{author}{\bibfnamefont{N.}~\bibnamefont{Ni}},
  \bibinfo{author}{\bibfnamefont{A.}~\bibnamefont{Thaler}},
  \bibinfo{author}{\bibfnamefont{S.~L.} \bibnamefont{Bud'ko}},
  \bibinfo{author}{\bibfnamefont{P.~C.} \bibnamefont{Canfield}},
  \bibnamefont{et~al.}, \bibinfo{journal}{Phys. Rev. B}
  \textbf{\bibinfo{volume}{81}}, \bibinfo{pages}{134512}
  (\bibinfo{year}{2010}).

\bibitem[{\citenamefont{Wang et~al.}(2010)\citenamefont{Wang, Luo, Zhao, Zhang,
  Wang, Marty, Chi, Lynn, Schneidewind, Li et~al.}}]{wang_electron-doping_2010}
\bibinfo{author}{\bibfnamefont{M.}~\bibnamefont{Wang}},
  \bibinfo{author}{\bibfnamefont{H.}~\bibnamefont{Luo}},
  \bibinfo{author}{\bibfnamefont{J.}~\bibnamefont{Zhao}},
  \bibinfo{author}{\bibfnamefont{C.}~\bibnamefont{Zhang}},
  \bibinfo{author}{\bibfnamefont{M.}~\bibnamefont{Wang}},
  \bibinfo{author}{\bibfnamefont{K.}~\bibnamefont{Marty}},
  \bibinfo{author}{\bibfnamefont{S.}~\bibnamefont{Chi}},
  \bibinfo{author}{\bibfnamefont{J.~W.} \bibnamefont{Lynn}},
  \bibinfo{author}{\bibfnamefont{A.}~\bibnamefont{Schneidewind}},
  \bibinfo{author}{\bibfnamefont{S.}~\bibnamefont{Li}}, \bibnamefont{et~al.},
  \bibinfo{journal}{Phys. Rev. B} \textbf{\bibinfo{volume}{81}},
  \bibinfo{pages}{174524} (\bibinfo{year}{2010}).

\bibitem[{\citenamefont{Nandi et~al.}(2010)\citenamefont{Nandi, Kim, Kreyssig,
  Fernandes, Pratt, Thaler, Ni, Bud'ko, Canfield, Schmalian
  et~al.}}]{nandi_anomalous_2010}
\bibinfo{author}{\bibfnamefont{S.}~\bibnamefont{Nandi}},
  \bibinfo{author}{\bibfnamefont{M.~G.} \bibnamefont{Kim}},
  \bibinfo{author}{\bibfnamefont{A.}~\bibnamefont{Kreyssig}},
  \bibinfo{author}{\bibfnamefont{R.~M.} \bibnamefont{Fernandes}},
  \bibinfo{author}{\bibfnamefont{D.~K.} \bibnamefont{Pratt}},
  \bibinfo{author}{\bibfnamefont{A.}~\bibnamefont{Thaler}},
  \bibinfo{author}{\bibfnamefont{N.}~\bibnamefont{Ni}},
  \bibinfo{author}{\bibfnamefont{S.~L.} \bibnamefont{Bud'ko}},
  \bibinfo{author}{\bibfnamefont{P.~C.} \bibnamefont{Canfield}},
  \bibinfo{author}{\bibfnamefont{J.}~\bibnamefont{Schmalian}},
  \bibnamefont{et~al.}, \bibinfo{journal}{Phys. Rev. Lett.}
  \textbf{\bibinfo{volume}{104}}, \bibinfo{pages}{057006}
  (\bibinfo{year}{2010}).

\bibitem[{\citenamefont{Fang et~al.}(2008)\citenamefont{Fang, Yao, Tsai, Hu,
  and Kivelson}}]{fang_theory_2008}
\bibinfo{author}{\bibfnamefont{C.}~\bibnamefont{Fang}},
  \bibinfo{author}{\bibfnamefont{H.}~\bibnamefont{Yao}},
  \bibinfo{author}{\bibfnamefont{W.-F.} \bibnamefont{Tsai}},
  \bibinfo{author}{\bibfnamefont{J.~P.} \bibnamefont{Hu}}, \bibnamefont{and}
  \bibinfo{author}{\bibfnamefont{S.~A.} \bibnamefont{Kivelson}},
  \bibinfo{journal}{Phys. Rev. B} \textbf{\bibinfo{volume}{77}},
  \bibinfo{pages}{224509} (\bibinfo{year}{2008}).

\bibitem[{\citenamefont{Xu et~al.}(2008)\citenamefont{Xu, M\"{u}ller, and
  Sachdev}}]{xu_ising_2008}
\bibinfo{author}{\bibfnamefont{C.}~\bibnamefont{Xu}},
  \bibinfo{author}{\bibfnamefont{M.}~\bibnamefont{M\"{u}ller}},
  \bibnamefont{and} \bibinfo{author}{\bibfnamefont{S.}~\bibnamefont{Sachdev}},
  \bibinfo{journal}{Phys. Rev. B} \textbf{\bibinfo{volume}{78}},
  \bibinfo{pages}{020501(R)} (\bibinfo{year}{2008}).

\bibitem[{\citenamefont{Sefat et~al.}(2009)\citenamefont{Sefat, Singh,
  VanBebber, Mozharivskyj, McGuire, Jin, Sales, Keppens, and
  Mandrus}}]{Sefat_2009}
\bibinfo{author}{\bibfnamefont{A.~S.} \bibnamefont{Sefat}},
  \bibinfo{author}{\bibfnamefont{D.~J.} \bibnamefont{Singh}},
  \bibinfo{author}{\bibfnamefont{L.~H.} \bibnamefont{VanBebber}},
  \bibinfo{author}{\bibfnamefont{Y.}~\bibnamefont{Mozharivskyj}},
  \bibinfo{author}{\bibfnamefont{M.~A.} \bibnamefont{McGuire}},
  \bibinfo{author}{\bibfnamefont{R.}~\bibnamefont{Jin}},
  \bibinfo{author}{\bibfnamefont{B.~C.} \bibnamefont{Sales}},
  \bibinfo{author}{\bibfnamefont{V.}~\bibnamefont{Keppens}}, \bibnamefont{and}
  \bibinfo{author}{\bibfnamefont{D.}~\bibnamefont{Mandrus}},
  \bibinfo{journal}{Phys. Rev. B} \textbf{\bibinfo{volume}{79}},
  \bibinfo{pages}{224524} (\bibinfo{year}{2009}).

\bibitem[{\citenamefont{Bud'ko et~al.}(2009)\citenamefont{Bud'ko, Nandi, Ni,
  Thaler, Kreyssig, Kracher, Yan, Goldman, and Canfield}}]{Budko_2009}
\bibinfo{author}{\bibfnamefont{S.~L.} \bibnamefont{Bud'ko}},
  \bibinfo{author}{\bibfnamefont{S.}~\bibnamefont{Nandi}},
  \bibinfo{author}{\bibfnamefont{N.}~\bibnamefont{Ni}},
  \bibinfo{author}{\bibfnamefont{A.}~\bibnamefont{Thaler}},
  \bibinfo{author}{\bibfnamefont{A.}~\bibnamefont{Kreyssig}},
  \bibinfo{author}{\bibfnamefont{A.}~\bibnamefont{Kracher}},
  \bibinfo{author}{\bibfnamefont{J.-Q.} \bibnamefont{Yan}},
  \bibinfo{author}{\bibfnamefont{A.~I.} \bibnamefont{Goldman}},
  \bibnamefont{and} \bibinfo{author}{\bibfnamefont{P.~C.}
  \bibnamefont{Canfield}}, \bibinfo{journal}{Phys. Rev. B}
  \textbf{\bibinfo{volume}{80}}, \bibinfo{pages}{014522}
  (\bibinfo{year}{2009}).

\bibitem[{\citenamefont{Marty et~al.}(2010)\citenamefont{Marty, Christianson,
  Wang, Matsuda, Cao, VanBebber, Zarestky, Singh, Sefat, and
  Lumsden}}]{Marty_2010}
\bibinfo{author}{\bibfnamefont{K.}~\bibnamefont{Marty}},
  \bibinfo{author}{\bibfnamefont{A.~D.} \bibnamefont{Christianson}},
  \bibinfo{author}{\bibfnamefont{C.~H.} \bibnamefont{Wang}},
  \bibinfo{author}{\bibfnamefont{M.}~\bibnamefont{Matsuda}},
  \bibinfo{author}{\bibfnamefont{H.}~\bibnamefont{Cao}},
  \bibinfo{author}{\bibfnamefont{L.~H.} \bibnamefont{VanBebber}},
  \bibinfo{author}{\bibfnamefont{J.~L.} \bibnamefont{Zarestky}},
  \bibinfo{author}{\bibfnamefont{D.~J.} \bibnamefont{Singh}},
  \bibinfo{author}{\bibfnamefont{A.~S.} \bibnamefont{Sefat}}, \bibnamefont{and}
  \bibinfo{author}{\bibfnamefont{M.~D.} \bibnamefont{Lumsden}},
  \bibinfo{journal}{arXiv:1009.1818}  (\bibinfo{year}{2010}),
  \bibinfo{note}{unpublished}.

\bibitem[{\citenamefont{Kim et~al.}(2010{\natexlab{a}})\citenamefont{Kim, Khim,
  Kim, Eom, Law, Kremer, Shim, and Kim}}]{Kim_2010}
\bibinfo{author}{\bibfnamefont{J.~S.} \bibnamefont{Kim}},
  \bibinfo{author}{\bibfnamefont{S.}~\bibnamefont{Khim}},
  \bibinfo{author}{\bibfnamefont{H.~J.} \bibnamefont{Kim}},
  \bibinfo{author}{\bibfnamefont{M.~J.} \bibnamefont{Eom}},
  \bibinfo{author}{\bibfnamefont{J.~M.} \bibnamefont{Law}},
  \bibinfo{author}{\bibfnamefont{R.~K.} \bibnamefont{Kremer}},
  \bibinfo{author}{\bibfnamefont{J.~H.} \bibnamefont{Shim}}, \bibnamefont{and}
  \bibinfo{author}{\bibfnamefont{K.~H.} \bibnamefont{Kim}},
  \bibinfo{journal}{Phys. Rev. B} \textbf{\bibinfo{volume}{82}},
  \bibinfo{pages}{024510} (\bibinfo{year}{2010}{\natexlab{a}}).

\bibitem[{\citenamefont{Liu et~al.}(2010)\citenamefont{Liu, Sun, Park, and
  Lin}}]{Liu_2010}
\bibinfo{author}{\bibfnamefont{Y.}~\bibnamefont{Liu}},
  \bibinfo{author}{\bibfnamefont{D.~L.} \bibnamefont{Sun}},
  \bibinfo{author}{\bibfnamefont{J.~T.} \bibnamefont{Park}}, \bibnamefont{and}
  \bibinfo{author}{\bibfnamefont{C.~T.} \bibnamefont{Lin}},
  \bibinfo{journal}{Physica C}  (\bibinfo{year}{2010}), \bibinfo{note}{in
  press}.

\bibitem[{\citenamefont{Kim et~al.}(2010{\natexlab{b}})\citenamefont{Kim,
  Kreyssig, Thaler, Pratt, Tian, Zarestky, Green, Bud'ko, Canfield, McQueeney
  et~al.}}]{mgkim_mn_2010}
\bibinfo{author}{\bibfnamefont{M.~G.} \bibnamefont{Kim}},
  \bibinfo{author}{\bibfnamefont{A.}~\bibnamefont{Kreyssig}},
  \bibinfo{author}{\bibfnamefont{A.}~\bibnamefont{Thaler}},
  \bibinfo{author}{\bibfnamefont{D.~K.} \bibnamefont{Pratt}},
  \bibinfo{author}{\bibfnamefont{W.}~\bibnamefont{Tian}},
  \bibinfo{author}{\bibfnamefont{J.~L.} \bibnamefont{Zarestky}},
  \bibinfo{author}{\bibfnamefont{M.~A.} \bibnamefont{Green}},
  \bibinfo{author}{\bibfnamefont{S.~L.} \bibnamefont{Bud'ko}},
  \bibinfo{author}{\bibfnamefont{P.~C.} \bibnamefont{Canfield}},
  \bibinfo{author}{\bibfnamefont{R.~J.} \bibnamefont{McQueeney}},
  \bibnamefont{et~al.}, \bibinfo{journal}{Phys. Rev. B}
  (\bibinfo{year}{2010}{\natexlab{b}}), \bibinfo{note}{in press,
  arXiv:1011.2816}.

\bibitem[{\citenamefont{Singh et~al.}(2009{\natexlab{a}})\citenamefont{Singh,
  Green, Huang, Kreyssig, McQueeney, Johnston, and Goldman}}]{YSingh_2009}
\bibinfo{author}{\bibfnamefont{Y.}~\bibnamefont{Singh}},
  \bibinfo{author}{\bibfnamefont{M.~A.} \bibnamefont{Green}},
  \bibinfo{author}{\bibfnamefont{Q.}~\bibnamefont{Huang}},
  \bibinfo{author}{\bibfnamefont{A.}~\bibnamefont{Kreyssig}},
  \bibinfo{author}{\bibfnamefont{R.~J.} \bibnamefont{McQueeney}},
  \bibinfo{author}{\bibfnamefont{D.~C.} \bibnamefont{Johnston}},
  \bibnamefont{and} \bibinfo{author}{\bibfnamefont{A.~I.}
  \bibnamefont{Goldman}}, \bibinfo{journal}{Phys. Rev. B}
  \textbf{\bibinfo{volume}{80}}, \bibinfo{pages}{100403(R)}
  (\bibinfo{year}{2009}{\natexlab{a}}).

\bibitem[{\citenamefont{Singh et~al.}(2009{\natexlab{b}})\citenamefont{Singh,
  Sefat, McGuire, Sales, Mandrus, VanBebber, and Keppens}}]{DSingh_2009}
\bibinfo{author}{\bibfnamefont{D.~J.} \bibnamefont{Singh}},
  \bibinfo{author}{\bibfnamefont{A.~S.} \bibnamefont{Sefat}},
  \bibinfo{author}{\bibfnamefont{M.~A.} \bibnamefont{McGuire}},
  \bibinfo{author}{\bibfnamefont{B.~C.} \bibnamefont{Sales}},
  \bibinfo{author}{\bibfnamefont{D.}~\bibnamefont{Mandrus}},
  \bibinfo{author}{\bibfnamefont{L.~H.} \bibnamefont{VanBebber}},
  \bibnamefont{and} \bibinfo{author}{\bibfnamefont{V.}~\bibnamefont{Keppens}},
  \bibinfo{journal}{Phys. Rev. B} \textbf{\bibinfo{volume}{79}},
  \bibinfo{pages}{094429} (\bibinfo{year}{2009}{\natexlab{b}}).

\bibitem[{\citenamefont{Jiang et~al.}(2009)\citenamefont{Jiang, Xing, Xuan,
  Wang, Ren, Feng, Dai, Xu, and Cao}}]{jiang_superconductivity_2009}
\bibinfo{author}{\bibfnamefont{S.}~\bibnamefont{Jiang}},
  \bibinfo{author}{\bibfnamefont{H.}~\bibnamefont{Xing}},
  \bibinfo{author}{\bibfnamefont{G.}~\bibnamefont{Xuan}},
  \bibinfo{author}{\bibfnamefont{C.}~\bibnamefont{Wang}},
  \bibinfo{author}{\bibfnamefont{Z.}~\bibnamefont{Ren}},
  \bibinfo{author}{\bibfnamefont{C.}~\bibnamefont{Feng}},
  \bibinfo{author}{\bibfnamefont{J.}~\bibnamefont{Dai}},
  \bibinfo{author}{\bibfnamefont{Z.}~\bibnamefont{Xu}}, \bibnamefont{and}
  \bibinfo{author}{\bibfnamefont{G.}~\bibnamefont{Cao}}, \bibinfo{journal}{J.
  Phys.: Condens. Matter} \textbf{\bibinfo{volume}{21}},
  \bibinfo{pages}{382203} (\bibinfo{year}{2009}).

\bibitem[{\citenamefont{Klintberg et~al.}(2010)\citenamefont{Klintberg, Goh,
  Kasahar, Nakai, Ishida, Sutherland, Shibauchi, Matsuda, and
  Terashima}}]{Klintberg_2010}
\bibinfo{author}{\bibfnamefont{L.~E.} \bibnamefont{Klintberg}},
  \bibinfo{author}{\bibfnamefont{S.~K.} \bibnamefont{Goh}},
  \bibinfo{author}{\bibfnamefont{S.}~\bibnamefont{Kasahar}},
  \bibinfo{author}{\bibfnamefont{Y.}~\bibnamefont{Nakai}},
  \bibinfo{author}{\bibfnamefont{K.}~\bibnamefont{Ishida}},
  \bibinfo{author}{\bibfnamefont{M.}~\bibnamefont{Sutherland}},
  \bibinfo{author}{\bibfnamefont{T.}~\bibnamefont{Shibauchi}},
  \bibinfo{author}{\bibfnamefont{Y.}~\bibnamefont{Matsuda}}, \bibnamefont{and}
  \bibinfo{author}{\bibfnamefont{T.}~\bibnamefont{Terashima}},
  \bibinfo{journal}{J. Phys. Soc. Jpn.} \textbf{\bibinfo{volume}{79}},
  \bibinfo{pages}{123706} (\bibinfo{year}{2010}).

\bibitem[{\citenamefont{Shishido et~al.}(2010)\citenamefont{Shishido, Bangura,
  Coldea, Tonegawa, Hashimoto, Kasahara, Rourke, Ikeda, Terashima, Settai
  et~al.}}]{shishido_evolution_2010}
\bibinfo{author}{\bibfnamefont{H.}~\bibnamefont{Shishido}},
  \bibinfo{author}{\bibfnamefont{A.~F.} \bibnamefont{Bangura}},
  \bibinfo{author}{\bibfnamefont{A.~I.} \bibnamefont{Coldea}},
  \bibinfo{author}{\bibfnamefont{S.}~\bibnamefont{Tonegawa}},
  \bibinfo{author}{\bibfnamefont{K.}~\bibnamefont{Hashimoto}},
  \bibinfo{author}{\bibfnamefont{S.}~\bibnamefont{Kasahara}},
  \bibinfo{author}{\bibfnamefont{P.~M.~C.} \bibnamefont{Rourke}},
  \bibinfo{author}{\bibfnamefont{H.}~\bibnamefont{Ikeda}},
  \bibinfo{author}{\bibfnamefont{T.}~\bibnamefont{Terashima}},
  \bibinfo{author}{\bibfnamefont{R.}~\bibnamefont{Settai}},
  \bibnamefont{et~al.}, \bibinfo{journal}{Phys. Rev. Lett.}
  \textbf{\bibinfo{volume}{104}}, \bibinfo{pages}{057008}
  (\bibinfo{year}{2010}).

\bibitem[{\citenamefont{Schnelle et~al.}(2009)\citenamefont{Schnelle,
  Leithe-Jasper, Gumeniuk, Burkhardt, Kasinathan, and Rosner}}]{Schnelle_2009}
\bibinfo{author}{\bibfnamefont{W.}~\bibnamefont{Schnelle}},
  \bibinfo{author}{\bibfnamefont{A.}~\bibnamefont{Leithe-Jasper}},
  \bibinfo{author}{\bibfnamefont{R.}~\bibnamefont{Gumeniuk}},
  \bibinfo{author}{\bibfnamefont{U.}~\bibnamefont{Burkhardt}},
  \bibinfo{author}{\bibfnamefont{D.}~\bibnamefont{Kasinathan}},
  \bibnamefont{and} \bibinfo{author}{\bibfnamefont{H.}~\bibnamefont{Rosner}},
  \bibinfo{journal}{Phys. Rev. B} \textbf{\bibinfo{volume}{79}},
  \bibinfo{pages}{214516} (\bibinfo{year}{2009}).

\bibitem[{\citenamefont{Qi et~al.}(2009)\citenamefont{Qi, Wang, Gao, Wang,
  Zhang, and Ma}}]{Qi_2009}
\bibinfo{author}{\bibfnamefont{Y.}~\bibnamefont{Qi}},
  \bibinfo{author}{\bibfnamefont{L.}~\bibnamefont{Wang}},
  \bibinfo{author}{\bibfnamefont{Z.}~\bibnamefont{Gao}},
  \bibinfo{author}{\bibfnamefont{W.}~\bibnamefont{Wang}},
  \bibinfo{author}{\bibfnamefont{X.}~\bibnamefont{Zhang}}, \bibnamefont{and}
  \bibinfo{author}{\bibfnamefont{Y.}~\bibnamefont{Ma}},
  \bibinfo{journal}{Physica C} \textbf{\bibinfo{volume}{469}},
  \bibinfo{pages}{1921} (\bibinfo{year}{2009}).

\bibitem[{\citenamefont{Sharma et~al.}(2010)\citenamefont{Sharma, Bharathi,
  Chandra, Reddy, Paulraj, Satya, Sastry, Gupta, and
  Sundar}}]{sharma_superconductivity_2010}
\bibinfo{author}{\bibfnamefont{S.}~\bibnamefont{Sharma}},
  \bibinfo{author}{\bibfnamefont{A.}~\bibnamefont{Bharathi}},
  \bibinfo{author}{\bibfnamefont{S.}~\bibnamefont{Chandra}},
  \bibinfo{author}{\bibfnamefont{V.~R.} \bibnamefont{Reddy}},
  \bibinfo{author}{\bibfnamefont{S.}~\bibnamefont{Paulraj}},
  \bibinfo{author}{\bibfnamefont{A.~T.} \bibnamefont{Satya}},
  \bibinfo{author}{\bibfnamefont{V.~S.} \bibnamefont{Sastry}},
  \bibinfo{author}{\bibfnamefont{A.}~\bibnamefont{Gupta}}, \bibnamefont{and}
  \bibinfo{author}{\bibfnamefont{C.~S.} \bibnamefont{Sundar}},
  \bibinfo{journal}{Phys. Rev. B} \textbf{\bibinfo{volume}{81}},
  \bibinfo{pages}{174512} (\bibinfo{year}{2010}).

\bibitem[{\citenamefont{{Rullier-Albenque}
  et~al.}(2010)\citenamefont{{Rullier-Albenque}, Colson, Forget, Thu\'{e}ry,
  and Poissonnet}}]{rullier-albenque_hole_2010}
\bibinfo{author}{\bibfnamefont{F.}~\bibnamefont{{Rullier-Albenque}}},
  \bibinfo{author}{\bibfnamefont{D.}~\bibnamefont{Colson}},
  \bibinfo{author}{\bibfnamefont{A.}~\bibnamefont{Forget}},
  \bibinfo{author}{\bibfnamefont{P.}~\bibnamefont{Thu\'{e}ry}},
  \bibnamefont{and}
  \bibinfo{author}{\bibfnamefont{S.}~\bibnamefont{Poissonnet}},
  \bibinfo{journal}{Phys. Rev. B} \textbf{\bibinfo{volume}{81}},
  \bibinfo{pages}{224503} (\bibinfo{year}{2010}).

\bibitem[{\citenamefont{Thaler et~al.}(2010)\citenamefont{Thaler, Ni, Kracher,
  Yan, Bud'ko, and Canfield}}]{thaler_physical_2010}
\bibinfo{author}{\bibfnamefont{A.}~\bibnamefont{Thaler}},
  \bibinfo{author}{\bibfnamefont{N.}~\bibnamefont{Ni}},
  \bibinfo{author}{\bibfnamefont{A.}~\bibnamefont{Kracher}},
  \bibinfo{author}{\bibfnamefont{J.~Q.} \bibnamefont{Yan}},
  \bibinfo{author}{\bibfnamefont{S.~L.} \bibnamefont{Bud'ko}},
  \bibnamefont{and} \bibinfo{author}{\bibfnamefont{P.~C.}
  \bibnamefont{Canfield}}, \bibinfo{journal}{Phys. Rev. B}
  \textbf{\bibinfo{volume}{82}}, \bibinfo{pages}{014534}
  (\bibinfo{year}{2010}).

\bibitem[{\citenamefont{Hodovanets et~al.}(2010)\citenamefont{Hodovanets, Mun,
  Thaler, Bud'ko, and Canfield}}]{hodovanets_2010}
\bibinfo{author}{\bibfnamefont{H.}~\bibnamefont{Hodovanets}},
  \bibinfo{author}{\bibfnamefont{E.~D.} \bibnamefont{Mun}},
  \bibinfo{author}{\bibfnamefont{A.}~\bibnamefont{Thaler}},
  \bibinfo{author}{\bibfnamefont{S.~L.} \bibnamefont{Bud'ko}},
  \bibnamefont{and} \bibinfo{author}{\bibfnamefont{P.~C.}
  \bibnamefont{Canfield}}, \bibinfo{journal}{arXiv:1010.5111}
  (\bibinfo{year}{2010}), \bibinfo{note}{unpublished}.

\bibitem[{\citenamefont{Kasahara et~al.}(2010)\citenamefont{Kasahara,
  Shibauchi, Hashimoto, Ikada, Tonegawa, Okazaki, Shishido, Ikeda, Takeya,
  Hirata et~al.}}]{Kasahara_2010}
\bibinfo{author}{\bibfnamefont{S.}~\bibnamefont{Kasahara}},
  \bibinfo{author}{\bibfnamefont{T.}~\bibnamefont{Shibauchi}},
  \bibinfo{author}{\bibfnamefont{K.}~\bibnamefont{Hashimoto}},
  \bibinfo{author}{\bibfnamefont{K.}~\bibnamefont{Ikada}},
  \bibinfo{author}{\bibfnamefont{S.}~\bibnamefont{Tonegawa}},
  \bibinfo{author}{\bibfnamefont{R.}~\bibnamefont{Okazaki}},
  \bibinfo{author}{\bibfnamefont{H.}~\bibnamefont{Shishido}},
  \bibinfo{author}{\bibfnamefont{H.}~\bibnamefont{Ikeda}},
  \bibinfo{author}{\bibfnamefont{H.}~\bibnamefont{Takeya}},
  \bibinfo{author}{\bibfnamefont{K.}~\bibnamefont{Hirata}},
  \bibnamefont{et~al.}, \bibinfo{journal}{Phys. Rev. B}
  \textbf{\bibinfo{volume}{81}}, \bibinfo{pages}{184519}
  (\bibinfo{year}{2010}).

\bibitem[{\citenamefont{Fernandes et~al.}(2010)\citenamefont{Fernandes, Pratt,
  Tian, Zarestky, Kreyssig, Nandi, Kim, Thaler, Ni, Canfield
  et~al.}}]{fernandes_unconventional_2010}
\bibinfo{author}{\bibfnamefont{R.~M.} \bibnamefont{Fernandes}},
  \bibinfo{author}{\bibfnamefont{D.~K.} \bibnamefont{Pratt}},
  \bibinfo{author}{\bibfnamefont{W.}~\bibnamefont{Tian}},
  \bibinfo{author}{\bibfnamefont{J.}~\bibnamefont{Zarestky}},
  \bibinfo{author}{\bibfnamefont{A.}~\bibnamefont{Kreyssig}},
  \bibinfo{author}{\bibfnamefont{S.}~\bibnamefont{Nandi}},
  \bibinfo{author}{\bibfnamefont{M.~G.} \bibnamefont{Kim}},
  \bibinfo{author}{\bibfnamefont{A.}~\bibnamefont{Thaler}},
  \bibinfo{author}{\bibfnamefont{N.}~\bibnamefont{Ni}},
  \bibinfo{author}{\bibfnamefont{P.~C.} \bibnamefont{Canfield}},
  \bibnamefont{et~al.}, \bibinfo{journal}{Phys. Rev. B}
  \textbf{\bibinfo{volume}{81}}, \bibinfo{pages}{140501(R)}
  (\bibinfo{year}{2010}).

\bibitem[{\citenamefont{Kimber et~al.}(2009)\citenamefont{Kimber, Kreyssig,
  Zhang, Jeschke, Valenti, Yokaichiya, Colombier, Yan, Hansen, Chatterji
  et~al.}}]{kimber_similarities_2009}
\bibinfo{author}{\bibfnamefont{S.~A.~J.} \bibnamefont{Kimber}},
  \bibinfo{author}{\bibfnamefont{A.}~\bibnamefont{Kreyssig}},
  \bibinfo{author}{\bibfnamefont{Y.-Z.} \bibnamefont{Zhang}},
  \bibinfo{author}{\bibfnamefont{H.~O.} \bibnamefont{Jeschke}},
  \bibinfo{author}{\bibfnamefont{R.}~\bibnamefont{Valenti}},
  \bibinfo{author}{\bibfnamefont{F.}~\bibnamefont{Yokaichiya}},
  \bibinfo{author}{\bibfnamefont{E.}~\bibnamefont{Colombier}},
  \bibinfo{author}{\bibfnamefont{J.}~\bibnamefont{Yan}},
  \bibinfo{author}{\bibfnamefont{T.~C.} \bibnamefont{Hansen}},
  \bibinfo{author}{\bibfnamefont{T.}~\bibnamefont{Chatterji}},
  \bibnamefont{et~al.}, \bibinfo{journal}{Nat. Mater.}
  \textbf{\bibinfo{volume}{8}}, \bibinfo{pages}{471} (\bibinfo{year}{2009}).

\bibitem[{\citenamefont{Brouet et~al.}(2010)\citenamefont{Brouet,
  Rullier-Albenque, Marsi, Mansart, Aichhorn, Biermann, Faure, Perfetti,
  Taleb-Ibrahimi, Le~F\`{e}vre et~al.}}]{Brouet_2010}
\bibinfo{author}{\bibfnamefont{V.}~\bibnamefont{Brouet}},
  \bibinfo{author}{\bibfnamefont{F.}~\bibnamefont{Rullier-Albenque}},
  \bibinfo{author}{\bibfnamefont{M.}~\bibnamefont{Marsi}},
  \bibinfo{author}{\bibfnamefont{B.}~\bibnamefont{Mansart}},
  \bibinfo{author}{\bibfnamefont{M.}~\bibnamefont{Aichhorn}},
  \bibinfo{author}{\bibfnamefont{S.}~\bibnamefont{Biermann}},
  \bibinfo{author}{\bibfnamefont{J.}~\bibnamefont{Faure}},
  \bibinfo{author}{\bibfnamefont{L.}~\bibnamefont{Perfetti}},
  \bibinfo{author}{\bibfnamefont{A.}~\bibnamefont{Taleb-Ibrahimi}},
  \bibinfo{author}{\bibfnamefont{P.}~\bibnamefont{Le~F\`{e}vre}},
  \bibnamefont{et~al.}, \bibinfo{journal}{Phys. Rev. Lett.}
  \textbf{\bibinfo{volume}{105}}, \bibinfo{pages}{087001}
  (\bibinfo{year}{2010}).

\end{thebibliography}

\end{document}